\documentclass[prb,twocolumn,showpacs,aps]{revtex4}
\usepackage{graphicx}
\usepackage{amsmath}

\begin{document}
\title{Enhanced transmission of optically thick metallic films at infrared
wavelengths }
\author{Dezhuan Han, Xin Li, Fengqin Wu,}
\author{Jian Zi}
\email{jzi@fudan.edu.cn}
\affiliation{Surface Physics Laboratory and Department of Physics, Fudan University,
Shanghai 200433, People's Republic of China}
\date{\today}

\begin{abstract}
For an optically thick metallic film, the transmission for both $s$- and $p$%
-polarized waves is extremely low. If the metallic film is coated on both
sides with a finite dielectric layer, light transmission for $p$-polarized
waves can be enhanced considerably. This enhancement is not related to
surface plasmon-polaritions. Instead, it is due to the interplay between
Fabry-Perot interference in the coated dielectric layer and the existence of
the Brewster angle at the dielectric/metallic interface. It is shown that
the coated metallic films can act as excellent polarizers at infrared
wavelengths.
\end{abstract}

\maketitle

Metal surfaces are highly reflective over a very wide range of wavelengths.
This is the reason why metals are commonly used as mirrors in our daily life
and in optical technologies as well. It is known that the transmission of an
optically thick metallic film is very low for wavelengths in the visible
range or below. However, for $p$-polarized waves transmission could be very
high when incident light excites the coupled surface plasmon-polaritons
(SPPs) on both sides of the metallic film.\cite{dra:85} SPPs are a kind of
electromagnetic excitations existing at the interface between a metal and a
dielectric medium.\cite{rae:88} For a flat metallic surface, $p$-polarized
incident waves cannot directly excite SPPs owing to the wavevector mismatch
between incident waves and SPPs. Normally, an attenuated total reflection
technique is adopted to generate evanescent waves which can excite SPPs.\cite%
{rae:88} Extraordinary transmission has been also found for a metallic film
perforated with subwavelength hole arrays\cite{ebb:98} or slits.\cite{por:99}
It is believed that the excitation and coupling of SPPs on both surfaces of
the metallic film play an important role in such an extraordinary
transmission.

In this letter, we report theoretically an enhanced transmission of $p$%
-polarized waves at infrared wavelengths for an optically thick metallic
film coated with a dielectric layer on both sides. It is found that the
enhanced transmission is not due to SPPs. Instead, it relies on the
interplay between Fabry-Perot interference in the dielectric layer and the
existence of a Brewster angle window at the dielectric/metallic interface.

In our numerical simulations, without loss of generality, metallic films are
assumed to be Ag. The dielectric constant of Ag is described by the Drude
model%
\begin{equation}
\varepsilon (\omega )=1-\frac{\omega _{p}^{2}}{\omega ^{2}+i\gamma \omega },
\end{equation}%
where $\omega _{p}$ is the plasma frequency and $\gamma $ is the parameter
related to the energy loss. For Ag, the parameters used are $\omega
_{p}=1.15\times 10^{16}$ rad/s and $\gamma =9.81\times 10^{13}$ rad/s, which
are obtained by fitting to the experimental data\cite{pal:85} at near and
mid infrared wavelengths.

\begin{figure}[b]
\centerline{\includegraphics[angle=0,width=7.cm]{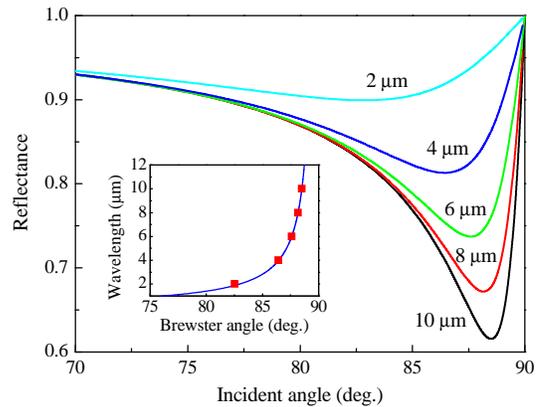}}
\caption{(color online). Reflectance spectra of $p$-polarized waves incident
from a dielectric medium (refractive index 1.5) upon a Ag surface at
different incident angles. Labels indicate different wavelengths. Inset
shows the calculated Brewster angles from Eq. (\protect\ref{bre}) and those
obtained from the reflectance spectra. }
\label{fig1}
\end{figure}

At a dielectric/dielectric interface, there exists a Brewster angle, at
which the reflectance for $p$-polarized waves is zero. To show the fact that
a metallic surface also possesses a Brewster angle, the reflectance spectra
for $p$-polarized waves incident from a dielectric medium with a refractive
index of $1.5$ upon a Ag surface at different incident angles are shown in
Fig. \ref{fig1}. For $s$-polarized waves, the reflection at the Ag surface
is very high and increases monotonically with increasing incident angle. For 
$p$-polarized waves, however, there exist dips in the reflectance spectra
for incident angles near 90$^{o}$. The positions of the dips can be viewed
as the Brewster angle at the Ag surface. Unlike the dielectric surface,
reflectance at the metallic surface is nonzero at the Brewster angle. At
short wavelengths (visible or near infrared), dips in the reflectance
spectra are rather shallow so that the Brewster angle is not well-defined.
Dips become sharper with the increase in wavelength. The Brewster angle is
wavelength dependent and shifts upward with the increasing wavelength. It
approaches 90$^{o}$ at the long wavelength limit. For a
dielectric/dielectric interface, the Brewster angle is given by $\tan \theta
_{B}=n_{2}/n_{1}$, where $n_{1}$ and $n_{2}$ are the refractive indices of
two dielectric media. For the dielectric/metallic interface, the Brewster
angle cannot be given by a simple formula.\cite{bed:01} However, for
wavelengths considered here it is found that the Brewster angle at the
dielectric/metallic interface can be well approximated by the following
relation%
\begin{equation}
\tan \theta _{B}=\left( \frac{\left\vert \varepsilon _{2}\right\vert }{%
\varepsilon _{1}}\right) ^{1/2}\text{,}  \label{bre}
\end{equation}%
where $\varepsilon _{1}$ and $\varepsilon _{2}$ are the dielectric constants
of the dielectric and metallic media, respectively. In Fig. \ref{fig1},
Brewster angles at different wavelengths obtained from the above relation
and from the reflectance spectra are also given for comparison. It can be
found that Brewster angles predicted by Eq. (\ref{bre}) agree well with
those obtained from the reflectance spectra.

\begin{figure}[t]
\centerline{\includegraphics[angle=0,width=7.cm]{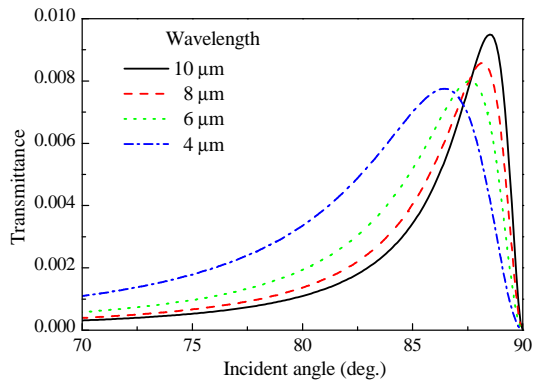}}
\caption{(color online). Transmittance spectra of $p$-polarized waves as for
a 80 nm Ag film immersed in a dielectric medium (refractive index 1.5) at
different incident angles. }
\label{fig2}
\end{figure}

The existence of the Brewster angle at a metallic surface can manifest a
higher transmission of $p$-polarized waves for a metallic film at the
Brewster incident angle with respect to other incident angles. Figure \ref%
{fig2} shows the transmittance spectra of $p$-polarized waves for a 80 nm Ag
film immersed in a dielectric medium at different incident angles. For $s$%
-polarized waves, transmission is extremely low for all incident angles.
Thus, the 80 nm Ag film is optically thick enough to block $s$-polarized
waves. For $p$-polarized waves, the situation, however, is a bit different.
The overall transmission is still very low, especially for incident angles
not close to grazing angles. It is interesting to note that there exist some
maxima in the transmittance spectra at some grazing incident angles close to
90$^{o}$. For a fixed wavelength transmittance increases monotonically for
incident angles varying from normal incidence, reaches a maximum at a
certain grazing incident angle, and then decreases monotonically up to the 90%
$^{o}$ incident angle. It is obvious that the positions of the transmission
peaks in Fig. \ref{fig2} coincide well with the Brewster angles at the
dielectric/metallic interface.

We now consider a Ag film coated with a dielectric layer symmetrically on
both sides. The refractive index of the coated dielectric layer is taken to
be 1.5. From the above discussions it is known that the Brewster angle is
close to 90$^{o}$. If $p$-polarized waves are incident from air upon the
dielectric layer, the refracted angle may be smaller than the Brewster angle
at the dielectric/metallic interface since the dielectric constant of air is
smaller than that of the dielectric medium. As a result, we may not access
the Brewster angle at the dielectric/metallic interface. To exclude this
possibility, two prisms with a higher refractive index than the coated
dielectric layer are introduced to situate on both coated dielectric layers.
Practically, it is not an easy job to get grazing incidence close to 90$^{0}$%
. The introduction of prisms, however, can overcome this difficulty. This
can be easily seen from Snell's law 
\begin{equation}
\sin \theta _{1}/\sin \theta _{2}=n_{d}/n_{p},
\end{equation}%
where $n_{p}$ and $n_{d}$ are the refractive indices of the prism and the
coated dielectric layer, respectively; $\theta _{1}$ is the incident angle
in the prism; $\theta _{2}$ is the refracted angle in the dielectric layer,
also the incident angle upon the metallic film. Without loss of generality,
the refractive index of prisms is chosen to be 3. Thus, the critical angle
at the prism/dielectric interface is 30$^{o}$. For $\theta _{1}$ larger than
this critical angle, evanescent waves will be generated. This configuration
has been commonly used to excite SPPs.\cite{rae:88}

\begin{figure}[b]
\centerline{\includegraphics[angle=0,width=7.cm]{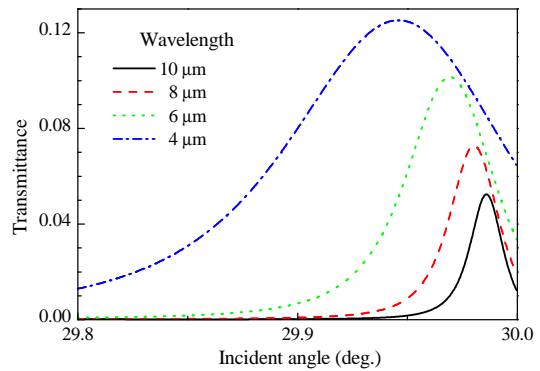}}
\caption{(color online). Transmittance spectra of $p$-polarized waves for a
coated Ag film with a thickness of 80 nm at different incident angles from
the prism. Labels indicate different wavelengths. The thickness of the
coated dielectric layer is determined from Eq. (\protect\ref{d-fp}). The
refractive indices of the coated dielectric and the prism are 1.5 and 3,
respectively.}
\label{fig3}
\end{figure}

In the coated layer, Fabry-Perot interference is expected owing to the
multiple reflection between the prism/dielectric and dielectric/metallic
interfaces. When traversing across the coated dielectric layer, two
successively transmitted waves have a phase difference 
\begin{equation}
\delta =\frac{2\pi }{\lambda _{0}}2n_{d}d\cos \theta _{2}+\phi ,
\end{equation}%
where $d$ is the thickness of the coated dielectric layer, $\lambda _{0}$ is
the wavelength in vacuum, and $\phi $ is an additional phase shift due to
the metallic surface. Transmission into the metallic film can be enhanced if 
$\delta =(2m+1)\pi $, where the integer $m$ takes the value of $0,1,2,...$.
Combining this effect with the Brewster angle window, enhanced transmission
should be expected.

In Fig. \ref{fig3} transmittance spectra for a coated 80 nm Ag film situated
between two prisms at difference incident angles are shown. To get enhanced
transmission the choice of the thickness of the coated dielectric layer is
crucial. At the Brewster angle, it is found that the additional phase
difference due to the metallic surface is $\pi /2$. Thus, the minimal
thickness of the coated dielectric layer that renders an enhanced
transmission is determined from%
\begin{equation}
d=\frac{\lambda _{0}}{8n_{d}\cos \theta _{B}}.  \label{d-fp}
\end{equation}%
It is obvious from Fig. \ref{fig3} that transmission is largely enhanced
with respect to the same metallic film situated in a dielectric medium. The
enhanced factor of the transmission peak for the wavelength of 4 $\mu $m is
over 16 and that for the wavelength of 10 $\mu $m is more than 5. It should
be noted that transmission is also dependent on the thickness of the
metallic film. Higher transmission can be obtained if we reduce the
thickness of the metallic film. For smaller thickness of the metallic film,
the minimal thickness of the coated dielectric layer that renders enhanced
transmission may deviate somewhat from that obtained from Eq. (\ref{d-fp})
owing to the coupling of the two dielectric/metallic interfaces.

\begin{figure}[t]
\centerline{\includegraphics[angle=0,width=7.5cm]{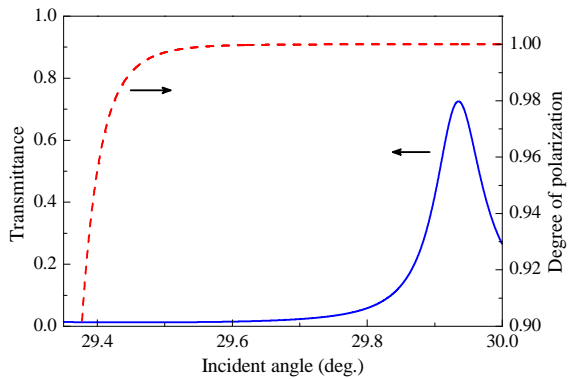}}
\caption{(color online). Transmittance spectrum of a $p$-polarized wave of $%
\protect\lambda _{0}=4$ $\protect\mu $m and degree of polarization for a
coated Ag film with a thickness of 40 nm as a function of the incident
angles from the prism. The thickness of the dielectric layer is 6.45 $%
\protect\mu $m.}
\label{fig4}
\end{figure}

As shown above, $p$-polarized waves possess a high transmission for a coated
metallic film, while $s$-polarized waves have an extremely low transmission.
This feature can render infrared polarizers with excellent performance
possible. To obtain a good performance, the thickness of the coated
dielectric layer and the metallic film should be chosen properly. A good
polarizer is characterized by a high transmission and a high degree of
polarization, defined by%
\begin{equation}
P=\frac{T_{p}-T_{s}}{T_{p}+T_{s}},
\end{equation}%
where $T_{p}$ and $T_{s}$ are the transmittance of $p$- and $s$-polarized
waves, respectively. For a perfect polarizer, the degree of polarization
should be $1$. In Fig. \ref{fig4}, the transmittance spectrum of a $p$%
-polarized wave with $\lambda _{0}=4$ $\mu $m for a coated Ag film with a
thickness of 40 nm and the degree of polarization of the system are shown. A
thickness of 6.45 $\mu $m for the dielectric layer is chosen in order to
obtain a maximal transmission at certain incident angle. It is obvious that
this polarizer has a high transmittance (over 70\%) for the $p$-polarized
wave around the incident angle of 29.93$^{o}$ from the prism, while its
degree of polarization is nearly perfect.

In the visible range, conventional polarizers have excellent performance and
are easily attainable. In the infrared regime, on the contrary, the
conventionally used metallic wire-type polarizers are costly and their
performance is less satisfactory with respect to those in the visible range.%
\cite{ben:78} Our results indicate that the coated metallic film could act
as an excellent polarizer at infrared wavelengths. The results shown in Fig. %
\ref{fig4} are for $\lambda _{0}=4$ $\mu $m. For other infrared wavelengths,
we can also obtain satisfactory performance for a coated metallic film
provided that the thicknesses of the coated dielectric layer and the
metallic film are properly chosen.

In summary, we studied theoretically the transmission of optically thick
metallic films. It was shown that there exists a Brewster angle window at
the dielectric/metallic interface. Incorporated with the Fabry-Perot
interference, a coated metallic film can have a largely enhanced
transmission. This feature can render excellent polarizers at infrared
wavelengths possible.

This work was supported by CNKBRSF, NSFC, PCSIRT, and Shanghai Science and
Technology Commission, China. We thank Dr. S. Meyer for a critical reading
of the manuscript.

\clearpage


\begin{thebibliography}{99}
\bibitem{dra:85} R. Dragila, B. Luther-Davies, and S. Vukovic, Phys. Rev.
Lett. \textbf{55}, 1117 (1985).

\bibitem{rae:88} H. Raether, \textit{Surface Plasmons }(Springer-Verlag,
Berlin, 1988).

\bibitem{ebb:98} T. W. Ebbesen, H. J. Lezec, H. F. Ghaemi, T. Thio, and P.
A. Wolff, Nature (London) \textbf{391}, 667 (1998).

\bibitem{por:99} J. A. Porto, F. J. Garc\'{\i}a-Vidal, and J. B. Pendry,
Phys. Rev. Lett. \textbf{83}, 2845 (1999).

\bibitem{pal:85} \textit{Handbook of Optical Constants of Solids, }edited by%
\textit{\ }E. D. Palik (Academic Press, New York, 1985).

\bibitem{bed:01} D. Bedeaux and J. Vlieger, \textit{Optical Properties of
Surfaces} (Imperial College Press, London, 2001), p. 64.

\bibitem{ben:78} J. M. Bennett and H. E. Bennett, \textit{Handbook of
Optics, }Ed. by W. Driscoll (McGraw-Hill, New York, 1978).
\end{thebibliography}
\end{document}